\documentclass[12pt]{iopart}
\usepackage{graphicx}
\usepackage{siunitx}
\usepackage{tensor}
\usepackage{amssymb}
\usepackage{booktabs} % For \toprule, \midrule, \bottomrule
\usepackage{multirow}
\usepackage{float}
\usepackage{xspace}
\usepackage[hidelinks]{hyperref}
\usepackage{subcaption} % Add subcaption package for subfigures
\newcommand{\antiproton}{\mbox{(anti-)proton}\xspace}
\newcommand{\antiprotons}{\mbox{(anti-)protons}\xspace}
\newcommand{\be}{$^9$Be$^+$\xspace}

\begin{document}

\title{A robust method to reach the motional quantum regime of \mbox{(anti-)protons} in cryogenic multi-Penning traps}

\author{Nikita Poljakov$^{1*}$, Jan Schaper$^1$, Julia-Aileen Coenders$^1$, Philipp Luca Hoffmann$^1$, \mbox{Juan Manuel Cornejo$^2$}, \mbox{Klemens Hammerer$^{3,4,5}$}, \mbox{Stefan Ulmer$^{6,7}$}, and \mbox{Christian Ospelkaus$^{1,8}$}}

\address{$^1$ Institut für Quantenoptik, Leibniz Universität Hannover, Germany}

\address{$^2$ Departamento de Física de la Materia Condensada, Universidad de Cádiz, Spain}

\address{$^3$ Institut fur Theoretische Physik, Leibniz Universität Hannover, Germany}

\address{$^4$ Institut für Theoretische Physik, Universität Innsbruck, Austria}

\address{$^5$ Institut für Quantenoptik und Quanteninformation, Österreichische Akademie der Wissenschaften, Austria}

\address{$^6$ Ulmer Fundamental Symmetries Laboratory, RIKEN, Japan}

\address{$^7$ Heinrich-Heine-Universität Düsseldorf, Germany}

\address{$^8$ Physikalisch-Technische Bundesanstalt, Braunschweig, Germany }

\ead{poljakov@iqo.uni-hannover.de}
\vspace{10pt}
\begin{indented}
\item[]February 2025
\end{indented}

\begin{abstract}

Sympathetic laser cooling is a key concept in precision spectroscopy and quantum state control of charged particles. Significant challenges arise in the metrologically relevant case where the effective interaction between the particles is weak and the particle to be cooled exhibits significant initial motional energy. Here we specifically address the most generally applicable case where the laser-cooled ion and the particle of interest are confined to two spatially separate potential wells  with equal motional frequency for resonant enhancement of the cooling dynamics. We analyze the latter through numerical simulations and find that anharmonicities of the potential wells can prevent maintaining the resonance condition throughout the cooling process and thus inhibit a significant reduction in motional energy. We propose a cooling scheme that sweeps the trapping frequency of the potential wells. We show that this scheme enables efficient cooling from cryogenic temperatures all the way to the quantum regime of motion. As a specific application scenario, we analyze the sympathetic cooling of \antiprotons into the quantum regime of motion for quantum-logic-spectroscopy-based tests of CPT invariance at the quantum limit in Penning traps. Nevertheless, our results and cooling strategies are generally applicable to other \mbox{laser-inaccessible} ion species. 

\end{abstract}

%
% Uncomment for keywords
\vspace{2pc}

%
% Uncomment for Submitted to journal title message
\submitto{Quantum Science and Technology}
%
% Uncomment if a separate title page is required
%\maketitle
% 
% For two-column output uncomment the next line and choose [10pt] rather than [12pt] in the \documentclass declaration
%\ioptwocol
%

\section{Introduction}

Full motional control of individual quantum systems, enabled through laser-cooling to the ground state of motion~\cite{diedrichLaserCoolingZeroPoint1989}, has opened new avenues for a wide range of applications in physics, ranging from quantum information processing~\cite{ciracQuantumComputationsCold1995, winelandExperimentalIssuesCoherent1998} to precision metrology~\cite{ludlowOpticalAtomicClocks2015}. The ability to coherently manipulate both internal and external degrees of freedom at the single-quantum level has enabled unprecedented levels of precision in such experiments~\cite{Marshall2025}. In the most general case, the particle of interest can however not be laser-cooled directly, either because of the lack of a suitable cooling transition or laser, or because direct cooling erases the internal quantum state. Sympathetic laser cooling~\cite{larsonSympatheticCoolingTrapped1986} and quantum logic spectroscopy (QLS)~\cite{heinzenQuantumlimitedCoolingDetection1990, schmidt_2005} address these issues by assigning coherent manipulation and spectroscopy to the ion of interest, and cooling and state detection to a ``cooling'' ion.

QLS, originally proposed for the \antiproton ~\cite{heinzenQuantumlimitedCoolingDetection1990,winelandExperimentalIssuesCoherent1998}, has first been demonstrated in the context of atomic clocks~\cite{schmidt_2005}, and later successfully applied to molecular~\cite{wolfNondestructiveStateDetection2016, chouPreparationCoherentManipulation2017} and highly charged ions~\cite{mickeCoherentLaserSpectroscopy2020}, all confined in radiofrequency traps. However, not all particles can be confined as Coulomb crystals in the same potential well as is the case for the aforementioned demonstrations. In particular, the antiproton cannot be confined in a single potential well together with the readily laser-cooled atomic cations. In the most general case, sympathetic cooling would therefore be implemented with the two particles in separate potential wells, connected either directly through their Coulomb interaction~\cite{winelandExperimentalIssuesCoherent1998} at close distance, or through induced image currents in a shared trap electrode~\cite{heinzenQuantumlimitedCoolingDetection1990}. Regarding the latter, the BASE collaboration~\cite{smorraBASEBaryonAntibaryon2015} was able to demonstrate sympathetic laser cooling of protons by a cloud of \be ions mediated through image currents in a shared electrode attached to a resonator~\cite{Will2024, bohmanSympatheticCoolingTrapped2021}, which can provide faster cooling and lower final temperatures than current Penning-trap cooling protocols~\cite{Latacz2025}. However, using that technique, it remains challenging to reach the quantum regime of motion in Penning traps. As to the former, although with both particles very close to the motional ground state, direct Coulomb coupling between separate potential wells has been demonstrated in radiofrequency traps~\cite{brownCoupledQuantizedMechanical2011,harlanderTrappedionAntennaeTransmission2011}. This method thus has the potential to achieve the quantum regime of motion and enable QLS in Penning traps~\cite{Cornejo_2021, Nitzschke2020}.

These advancements are particularly relevant for high-precision experiments, such as those conducted by the BASE collaboration in their search for discrepancies between protons and antiprotons to test the CPT (Charge-Parity-Time reversal) theorem~\cite{lehnertCPTSymmetryIts2016}. BASE is performing high-precision measurements of the charge-to-mass ratio~\cite{Borchert2022} and $g$-factors of the \antiproton~\cite{schneiderDoubletrapMeasurementProton2017, Smorra2017} in advanced cryogenic multi-Penning traps, and recently demonstrated coherent spin spectroscopy of an antiproton~\cite{Latacz2025}, enabling further improvements in the $g$-factor precision. A possible route for improving both precision and sampling rates in such experiments could be achieved by applying QLS techniques to Penning traps, where the preparation of the \antiproton in the motional ground state via sympathetic cooling is a key requisite. Ultimately, extending these techniques to a wider range of ion species could have a significant impact on high-precision experiments in Penning traps.

In this paper, we analyze the dynamics of sympathetic laser cooling in a realistic double-well potential, focusing on the scenario of cooling of an \antiproton through coupling to a laser-cooled \be ion. An implementation in any metrologically relevant scenario will, however, have to take into account that the particle of interest that cannot be directly laser-cooled may have a comparatively large initial axial motional energy after thermalizing with the cryogenic environment at \SI{4}{\kelvin}. To address that, we analyze the energy range of the harmonic region of the trapping potential and quantify the stability of energy transfer against voltage fluctuations. We show that, due to trap anharmonicities and limited voltage stability, sympathetic cooling is inefficient for higher-energy particles in the static double-well potential. We therefore propose a frequency-sweeping scheme that exploits these anharmonicities to enhance cooling. 

The manuscript is organized as follows. The ``Experimental setup'' section describes a multi-Penning trap stack and measurement techniques that can be implemented for the \antiproton-\be coupling. The ``Method'' section introduces the mathematical and numerical framework for simulating the trapping potentials and coupling dynamics. The ``Results and Discussion'' section presents the simulation results, including the ground-state cooling protocol. Finally, the ``Conclusion'' section provides a summary and an outlook. 

\section{Experimental setup}

\begin{figure}[b]
    \centering
    \includegraphics[width=\textwidth]{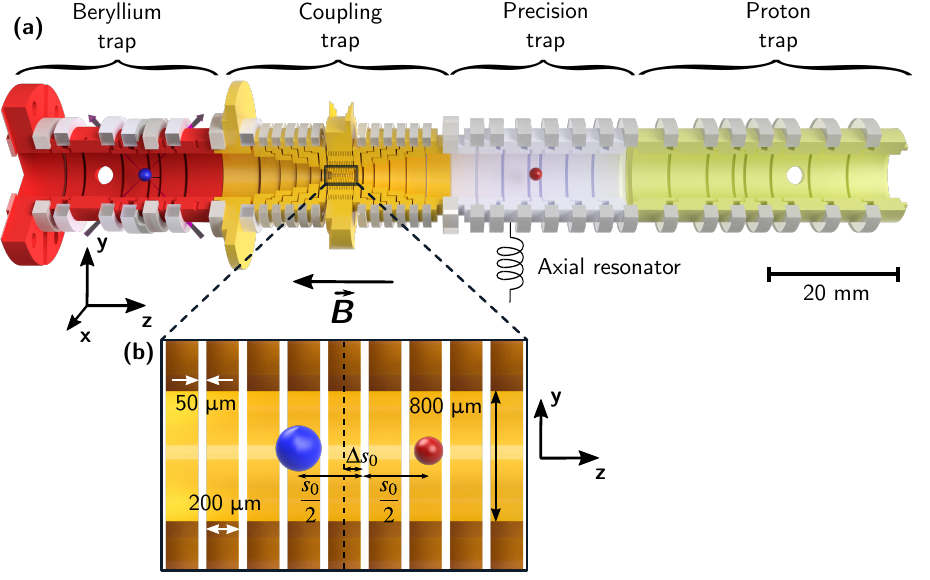}
    \caption{\textbf{Multi-Penning trap.} \textbf{a)} Cross-sectional view of an exemplary multi-Penning trap design for the coupling experiments, with trap sections shown in different colors. A \be ion is depicted in blue and an (anti-)proton in red. The inner trap surface is circularly symmetric around the trap axis. The macroscopic trap electrodes have inner diameters of \SI{9}{mm} and are made of gold-plated, oxygen-free high thermal conductivity copper. An axial resonator, consisting of a superconducting coil and a cryogenic amplifier, is connected to a precision trap electrode for \antiproton detection and \SI{4}{\kelvin} cooling. Purple arrows illustrate the Raman beams used for cooling and temperature measurement of the \be ion. \textbf{b)} Detailed cross-sectional view of the Coupling trap, consisting of nine microfabricated electrodes. The positions of the \antiproton and \be ion are shown, along with the particle separation $s_0$ and offset $\Delta s_0$ relative to the central electrode. Further details are provided in the text.}
    \label{fig:trapstack}
\end{figure}

A Penning trap confines charged particles using a combination of a strong, homogeneous magnetic field $\vec{B}$ in the axial direction and a static electric quadrupole potential. The trapped particle exhibits three independent harmonic modes of motion: two radial modes — the modified cyclotron and magnetron motions with frequencies $\omega_+$ and $\omega_-$, respectively — perpendicular to the magnetic field, and an axial motion parallel to it with a frequency $\omega_z$. These eigenmotions can be related to the free cyclotron frequency $\omega_c$ by the invariance theorem, $\omega_c=\sqrt{\omega_+^2+\omega_-^2+\omega_z^2}=q|\vec{B}|/m$~\cite{brownPrecisionSpectroscopyCharged1982}, where $q$ is the charge, $B$ the magnetic field strength, and $m$ the particle mass.

Fig.~\ref{fig:trapstack}a shows a schematic of a multi-Penning trap in which the sympathetic cooling scheme of the \antiproton via direct Coulomb interaction could be implemented. The trap stack consists of different trapping regions denoted with different colors. In our setup~\cite{niemannCryogenic9BePenning2019, mielke139GHzUV2021, cornejoOpticalStimulatedRamanSideband2023}, we employ \be as the cooling ion, as it is the laser-accessible species with the closest mass to the \antiproton. In the \mbox{``Beryllium trap''} (shown in red), a single \be ion is loaded using a pulsed \SI{532}{\nano\meter} ablation laser and a \SI{235}{\nano\meter} photoionization laser, and is cooled to the motional ground state using a resolved Raman sideband cooling scheme~\cite{cornejo_resolved-sideband_2024}. It has been shown that a proton can be reliably loaded using the same ablation laser on a tantalum target~\cite{VELARDI201420, Wiesinger2023}, which is placed in the \mbox{``Proton trap''} (green). The proton will be detected and cooled to \SI{4}{\kelvin} using an axial resonator in the \mbox{``Precision trap''} (white)~\cite{Nagahama2016}. An antiproton could be injected from a transportable Penning trap system of the \mbox{BASE-STEP} experiment at CERN~\cite{Leonhardt2025}. Using fast adiabatic transport~\cite{Meiners2024, vBoehn2025}, both \be and an \antiproton could be transported into the desired double-well potential of a microfabricated trap section, called the \mbox{``Coupling trap''}~\cite{niemannCryogenic9BePenning2019} (yellow) (see Fig.~\ref{fig:trapstack}b), where our ground-state cooling method of the \antiproton will be employed. After the coupling, the \be ion can be transported into the Beryllium trap to measure its temperature and to re-initialize it in the ground state, while the \antiproton can be stored on the other side of the Coupling trap. 

\newpage

The Coupling trap has an inner diameter of \SI{800}{\micro\meter} and is composed of \SI{200}{\micro\meter}-thick electrodes separated by \SI{50}{\micro\meter}. The electrodes are structured through selective laser-induced etching of a fused-silica substrate, followed by lithographic processing, and completed by electroplating to deposit the gold layer. The remaining unplated fused silica forms a C-shaped support structure that enables self-aligning stacking of the electrodes. The fabrication of the coupling trap is discussed in detail in~\cite{Julia2026}. The trap comprises nine electrodes, providing nine independent degrees of freedom. Its compact geometry reduces the constraints associated with shaping the double-well potential, as discussed in the following section.

\section{Method}

To simulate double-well potentials for the sympathetic cooling of the \antiproton, we first model the Coupling trap electrodes described above in \mbox{COMSOL}~\cite{comsol}. Using the finite element method, we obtain electrostatic potentials $\phi_i(z)$ and
electric fields $\frac{\partial \phi_i}{\partial z}(z)$ along the trap axis $z$ for each electrode $i$ individually set to \SI{1}{\volt}, and other electrodes $\neq i$ set to \SI{0}{\volt}. These values are then used to solve the following matrix equation~\cite{blakestadNeargroundstateTransportTrappedion2011,Meiners2024} for the voltages \mbox{$V_1$, $V_2$, ..., $V_n$} applied to $n=9$ electrodes,

\begin{equation}
\renewcommand{\arraystretch}{1.2}
\left(\begin{array}{cccc}
\frac{\partial \phi_1}{\partial z}|_{z=z_{a,0}} & \frac{\partial \phi_2}{\partial z}|_{z=z_{a,0}} & \ldots & \frac{\partial \phi_n}{\partial z}|_{z=z_{a,0}} \\
\frac{\partial \phi_1}{\partial z}|_{z=z_{b,0}} & \frac{\partial \phi_2}{\partial z}|_{z=z_{b,0}} & \ldots & \frac{\partial \phi_n}{\partial z}|_{z=z_{b,0}} \\
\frac{\partial^2 \phi_1}{\partial z^2}|_{z=z_{a,0}} & \frac{\partial^2 \phi_2}{\partial z^2}|_{z=z_{a,0}} & \ldots & \frac{\partial^2 \phi_n}{\partial z^2}|_{z=z_{a,0}} \\
\frac{\partial^2 \phi_1}{\partial z^2}|_{z=z_{b,0}} & \frac{\partial^2 \phi_2}{\partial z^2}|_{z=z_{b,0}} & \ldots & \frac{\partial^2 \phi_n}{\partial z^2}|_{z=z_{b,0}} \\
\end{array}\right)\left(\begin{array}{c}
V_1 \\
V_2 \\
\vdots \\
V_n
\end{array}\right)=\left(\begin{array}{c}
0 \\
0 \\
\frac{m_a \omega_a^2}{q_a} \\
\frac{m_b \omega_b^2}{q_b}
\end{array}\right),
\label{eqn:matrix}
\end{equation}
where $z_{a,0}$ and $z_{b,0}$ denote the positions of the double-well potential minima for charged particles $a$ and $b$, $m_a$ and $m_b$ are their masses, $q_a$ and $q_b$ are their charges, $\omega_a$ and $\omega_{b}$ are their angular oscillation frequencies. To minimize anharmonicities, we set the third- and fourth-order derivatives of the potential at its minima to zero; these terms are omitted in Eq.~\ref{eqn:matrix} for brevity. The solution with the smallest voltages is obtained using the Moore-Penrose inverse~\cite{penrose1955}. Using the superposition principle, the resulting voltages are combined to construct the double-well potential,

\begin{equation}
    \Phi(z) = \sum_{i=1}^n V_i\phi_i(z).
\end{equation}

\begin{figure}[t]
    \centering
    \includegraphics[width=\linewidth]{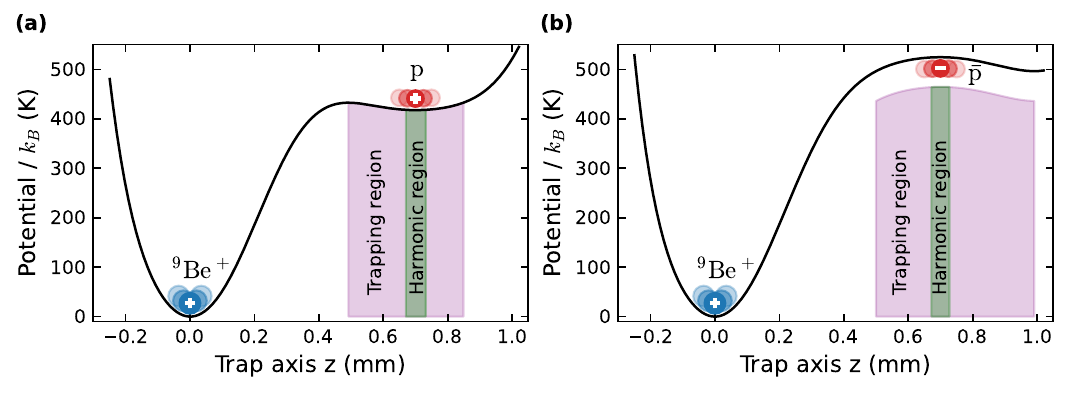}
    \caption{\textbf{Double-well potential.} \textbf{a)} Proton coupled with a \be ion. \textbf{b)} Antiproton coupled with a \be ion. The double-well potentials were generated using potential minima separation $s_0=$ \SI{0.7}{\milli\meter} and axial trap frequency $f=\omega/2\pi=$ \SI{500}{\kilo\hertz} for both potential minima. The antiproton potential well is inverted due to its negative charge. The purple shaded region indicates the positions where the \antiproton can be trapped, while the green shaded region represents positions where it experiences a harmonic potential. The energy ranges of the trapping and harmonic regions are described in more detail in Fig.~\ref{fig:harmoniccoupling}.}
    \label{fig:potentialshape}
\end{figure}

Examples of the resulting double-well potentials are shown in Fig.~\ref{fig:potentialshape}a for the proton and in Fig.~\ref{fig:potentialshape}b for the antiproton. The potentials are asymmetric due to the 9:1 mass ratio of the \be ion and the \antiproton. For the proton, the well depth is determined by the barrier height between the proton and the \be ion. For the antiproton, which carries a negative charge, there is no barrier between it and the \be ion; however, a finite barrier exists on the opposite side of the well. 

We simulate sympathetic cooling of the \antiproton in the double-well potential using the Verlet method~\cite{verlet1967}, which is symplectic and thus provides good energy conservation. The simulation timestep is chosen in the range of \SIrange{10}{50}{\nano\second}, which oversamples the axial oscillation period by at least a factor of 100 to ensure accurate energy conservation. At every timestep, we evaluate the force that particle $a$ experiences due to the Coulomb force from particle $b$ and electrostatic force of the trap,

\begin{equation}
    F_a(z_a,z_b) = \frac{1}{4\pi\epsilon_0}\frac{q_aq_b}{(z_a-z_b)^2} + q_a \left.\frac{\partial\Phi}{\partial z}\right|_{z=z_{a}},
\end{equation}
where $z_a$ and $z_b$ are particle positions, $\epsilon_0$ is the vacuum permittivity, and $\left.\frac{\partial\Phi}{\partial z}\right|_{z=z_{a}}$ is the electric field of the double-well potential experienced by the particle $a$. An analogous calculation is performed for particle $b$. Due to the second-order term of the Taylor expansion of the Coulomb potential~\cite{brownCoupledQuantizedMechanical2011}, the charged particles can be described as coupled harmonic oscillators behaving like oscillating masses connected by a spring if the resonance condition $\omega_a\approx\omega_b$ is satisfied. Therefore, as the particles oscillate in their respective potential wells, their energies are exchanged after the exchange time~\cite{brownCoupledQuantizedMechanical2011},

\begin{equation}
    \tau_{ex} = \frac{2\pi^2 \epsilon_0 s_0^3 \sqrt{m_a m_b} \sqrt{\omega_a \omega_b}}{q_a q_b}.
\label{eqn:transfer_time}
\end{equation}

In our simulations, the total energy of the particles during the exchange process is evaluated as the sum of their kinetic and electrostatic potential energies along the trap axis $z$. We consider only the axial motion, because the radial amplitude of the \antiproton can be reduced to $\sim$ \SI{1}{\micro\meter} by coupling the radial and axial motions using radiofrequency pulses applied to the trap electrodes~\cite{cornell1990, Brown1986}. For the \be ion, the radial amplitude can be reduced to \mbox{$\scriptstyle{\lesssim}$ \SI{1}{\micro\meter}} using laser cooling and axialization techniques~\cite{Powell2003}. Consequently, for both particles, the radial amplitude is negligible compared to $s_0$.

\begin{figure}[b]
    \centering
    \includegraphics[width=\textwidth]{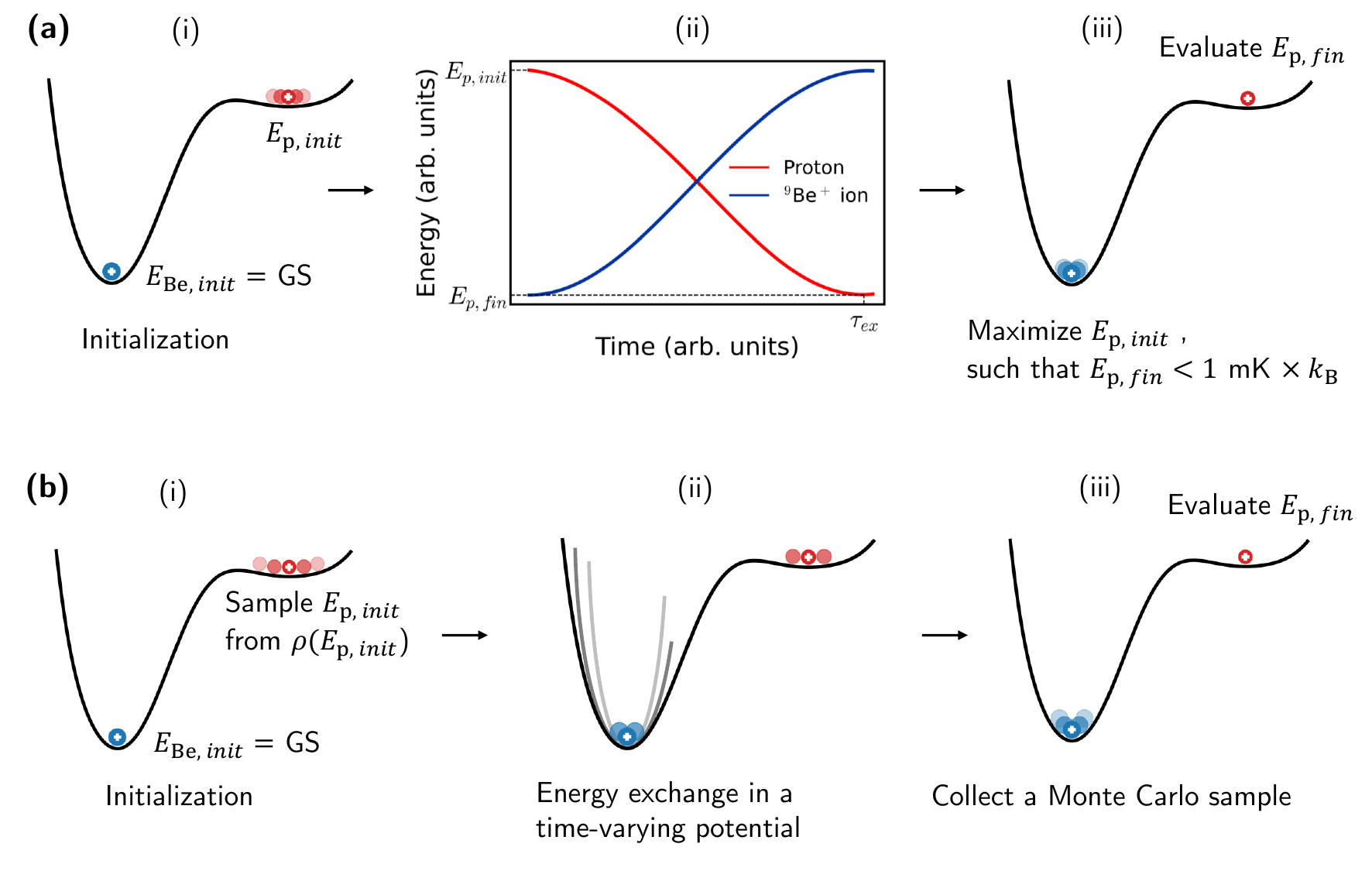}
    \caption{\textbf{Simulation sequence.} \textbf{(a)} Harmonic coupling. \textbf{(b)} Frequency sweep. The simulation sequences are shown using a proton as an example; the same studies are performed for the antiproton. Both subfigures are divided into three steps: i) initialization of the \antiproton with energy $E_{\rm{p/\bar{p}}, init}$ and preparation of the \be ion in its motional ground state (GS); ii) energy exchange between the particles; iii) analysis of the final \antiproton energy $E_{\rm{p/\bar{p}},fin}$. The plot in subfigure (a.ii) shows an example of the energy evolution of the \antiproton and \be ion during the coupling process in the time-independent double-well potential. }
    \label{fig:simulation_sequence}
\end{figure}

In this work, we present two numerical studies: 1) harmonic coupling (see subsection \ref{subsec:harmonic}) and 2) frequency sweep (see subsection \ref{subsec:sweep}). The numerical sequences of both studies are illustrated in Fig.~\ref{fig:simulation_sequence}. For the harmonic coupling study, we consider a time-independent double-well potential in which both particles are confined to harmonic regions around their respective potential minima. The \be ion is initialized in its motional ground state, while the \antiproton is prepared with an initial energy $E_{\rm{p/\bar{p}},init}$. During the coupling process, the \be ion is not subjected to laser cooling. From the second-order term of the Taylor expansion of the Coulomb potential, it follows that if $m_a \neq m_b$, the interaction induces an angular frequency shift $\Delta \omega$ between the two oscillating particles,

\begin{equation}
\Delta \omega =\left(\frac{1}{m_a}-\frac{1}{m_b}\right)\frac{q_a q_b}{4\pi\epsilon_0 \omega s_0^3}.
\label{eqn:freq_shift}
\end{equation}
The derivation of this formula is presented in~\ref{sec:freq_shift}. To maintain the resonance condition, the curvatures of the potential wells near their respective minima are chosen such that the oscillation frequencies of the two trapped particles are detuned by $\Delta\omega$. The simulation is run for at least the exchange time $\tau_{ex}$, during which the particle energy evolution is calculated. An example of this evolution is shown in Fig.~\ref{fig:simulation_sequence}a and exhibits the sinusoidal behavior expected for coupled harmonic oscillators. Finally, the minimum \antiproton energy $E_{\rm{p/\bar{p}},fin}$ is analyzed. The procedure is repeated while increasing $E_{\rm{p/\bar{p}},init}$ until the condition $E_{\rm{p/\bar{p}},fin}<$ \SI{1}{\milli\kelvin}$\times k_{\mathrm{B}}$ is no longer satisfied. In this way, the boundary of the harmonic region accessible to the \antiproton is estimated.

For the frequency sweep study, we consider a time-dependent potential, where the \antiproton is not necessarily confined to the harmonic region near its potential minimum, as it may possess a larger initial energy $E_{\rm{p/\bar{p}},init}$ than in the harmonic coupling study. To calculate the \mbox{(anti-)proton's} initial axial energy, $E_{\rm{p/\bar{p}},init}$, we model it as a canonical ensemble in thermal contact with the axial resonator coil, which acts as a thermal reservoir at \mbox{$T_z=$ \SI{4}{\kelvin}}. Accordingly, the \antiproton is initialized by taking random samples from the Boltzmann distribution, given by

\begin{equation}
    \rho\left(E_{\rm{p/\bar{p},init}} \right) = \frac{1}{k_B T_z} \exp{\left[-E_{\rm{p/\bar{p},init}}/(k_B T_z)\right]},
\label{eqn:boltzmann}
\end{equation}
where $k_B$ is the Boltzmann constant. The \be ion is again initialized in its motional ground state. We select a set of \be ion trapping frequencies for the sweep, and find the corresponding voltages by solving Eq.~\ref{eqn:matrix}. To obtain the intermediate frequency values, we perform linear interpolation of the voltages. During the frequency sweep, the energy is exchanged between the particles, and the final \antiproton energy $E_{\rm{p/\bar{p}}, fin}$ is evaluated at the last timestep of the sweep and recorded as a Monte Carlo sample. Further details about the numerical studies are discussed in the following section.

\section{Results and Discussion}

Here, we present the results of harmonic coupling and frequency sweep simulations which are subsequently combined to form a sympathetic cooling scheme.

\subsection{Harmonic coupling}
\label{subsec:harmonic}

We study the cooling dynamics of the \be-\antiproton coupling in time-independent double-well trapping potentials that can be produced in the Coupling trap using different trapping parameters. As shown in Tab.~\ref{tab:offsets}, we optimize the positions of particles with respect to the electrodes by introducing an axial offset $\Delta s_0$, such that it maximizes the energy that can be cooled using the harmonic region of the \antiproton potential. 

\begin{table}[h]
    \centering
    \caption{Optimized offsets, $\Delta s_{0,p}$ and $\Delta s_{0,\bar{p}}$, corresponding to the maximum harmonicity of the potential for the proton and antiproton, respectively. The offset is defined as the distance between the central electrode and the midpoint between the particles, as illustrated in Fig.~\ref{fig:trapstack}b.}
    \label{tab:offsets}
    \begin{tabular}{l ccc}
        \toprule
        \toprule
      $s_0$ (mm) & $\Delta s_{0,p}$ (\SI{}{\micro\meter}) & $\Delta s_{0,\bar{p}}$ (\SI{}{\micro\meter})\\
        \midrule
    0.6 & 50 & 50  \\
         0.7 & 32 & 53\\
         0.8 & 95 & 88 \\
        \bottomrule
        \bottomrule
    \end{tabular}
\end{table}

\begin{figure}[h]
    \centering
    \begin{subfigure}[b]{\textwidth}
        \centering
        \includegraphics[width=\textwidth]{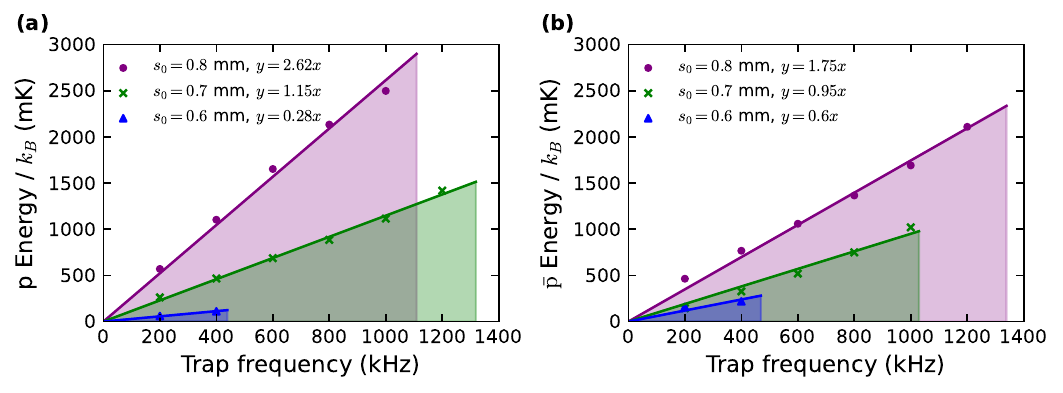}
    \end{subfigure}
    \begin{subfigure}[b]{\textwidth}
        \centering
        \includegraphics[width=\textwidth]{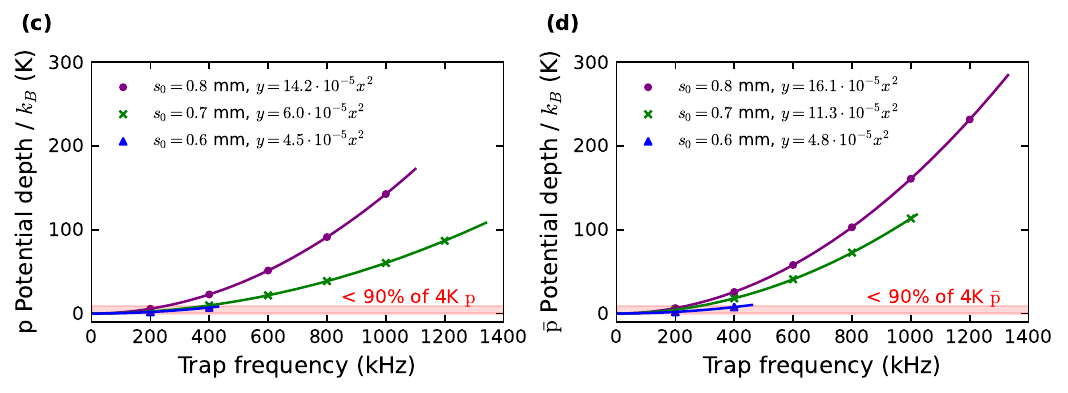}
    \end{subfigure}
    \caption{\textbf{Energy ranges of the double-well potential.} Harmonic energy ranges for the \textbf{(a)} proton and \textbf{(b)} antiproton, and trapping energy ranges for the \textbf{(c)} proton and \textbf{(d)} antiproton, as functions of trap frequency.
    Shaded regions in (a) and (b) denote initial energies from which the antiproton can be cooled below \(\SI{1}{\milli\kelvin}\,\times k_\mathrm{B}\) via harmonic coupling; their upper boundaries are given by linear fits (shown in the legend), with coefficients in \(\SI{}{\milli\kelvin\per\kilo\hertz}\). In (c) and (d), the data are fitted quadratically (shown in the legend), with coefficients in \(\SI{}{\milli\kelvin\per\kilo\hertz\squared}\). Potentials with parameters \(\omega\) and \(s_0\) above the red-shaded region trap more than 90\% of \(\SI{4}{\kelvin}\) (anti-)protons.}
    \label{fig:harmoniccoupling}
\end{figure}

\begin{figure}[t]
    \centering
    \includegraphics[width=\linewidth]{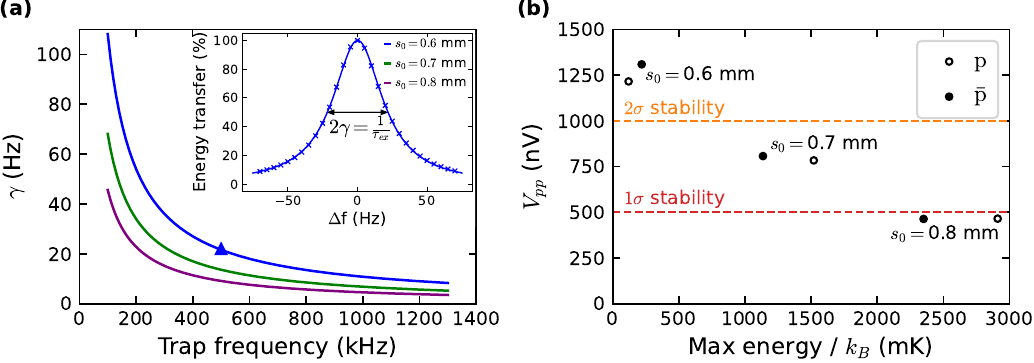}
    \caption{\textbf{Stability of the energy transfer.} \textbf{a)} Half width at half maximum ($\gamma$) of the energy transfer resonance curve as a function of the trap frequency. The plot was calculated using Eq.~\ref{eqn:transfer_time} and the relation $2\gamma=1/\tau_{ex}$. In the top right corner, an example resonance curve for $f=$ \SI{500}{\kilo\hertz} and $s_0=$ \SI{0.6}{\milli\meter} (marked with a blue triangle) is fit to a Lorentzian function. The data are evaluated using the percentage of energy transferred from a proton oscillating at frequency $f$ to a \be ion oscillating at frequency $f+\Delta f$, where $\Delta f$ is the deviation from the resonance frequency. \mbox{\textbf{b)} Required} power supply stability to achieve an 80\% energy transfer. For each $s_0$, the maximum initial energy of the \antiproton is taken from Fig.~\ref{fig:harmoniccoupling}. The calculation of the voltage stability is described in the appendix. Dashed lines indicate a power supply stability of \mbox{$V_{pp}=$ \SI{0.5}{\micro\volt_{pp}}}. Data points above the $1\sigma$ stability line are expected to achieve the desired transfer at least 68.3\% of the time, while those above the $2\sigma$ stability line are expected to do so at least 95.4\% of the time.}
    \label{fig:stability}
\end{figure}

We estimate the maximum energy from which an \antiproton can be cooled to below \SI{1}{\milli\kelvin}, as shown in Figs.~\ref{fig:harmoniccoupling}a and~\ref{fig:harmoniccoupling}b. In our analysis, we restrict the parameters to those achievable with electrode voltages below \SI{10}{\volt}, which sets an upper limit on the axial trap frequency $f=\omega/2\pi$ for each value of $s_0$ in the plots. Increasing both $f$ and $s_0$ allows to cool an \antiproton from a higher initial energy. Figs.~\ref{fig:harmoniccoupling}c and~\ref{fig:harmoniccoupling}d show the corresponding potential well depths, which scale quadratically with $f$. Consequently, larger harmonic regions are associated with deeper wells. While this suggests that larger $s_0$ and $f$ are advantageous, Fig.~\ref{fig:stability}a demonstrates that these conditions also result in a narrower resonance curve, imposing stricter requirements on frequency stability in the experimental implementation.

We investigate the voltage stability required to achieve an 80\% energy transfer, as shown in Fig.~\ref{fig:stability}b. The detailed calculations are presented in the appendix. The required stability is compared to that of a power supply with \mbox{$V_{pp}=$ \SI{0.5}{\micro\volt_{pp}}}, as estimated in the datasheet of the Stefan Stahl UM 1-14-LN ultrastable power supply~\cite{stahl_manual}. Using a power supply with larger peak-to-peak voltage fluctuations results in a reduced fraction of transferred energy. Cooling an \antiproton from a higher initial energy requires a larger $s_0$, which in turn again requires a higher power supply stability. Therefore, to cool the majority of the \antiproton energy distribution at \SI{4}{\kelvin} while relaxing the stability requirements, we consider a smaller $s_0$ in combination with a time-varying anharmonic potential.

\subsection{Frequency sweep}
\label{subsec:sweep}

\begin{figure}[h]
    \centering
    \includegraphics[width=\textwidth]{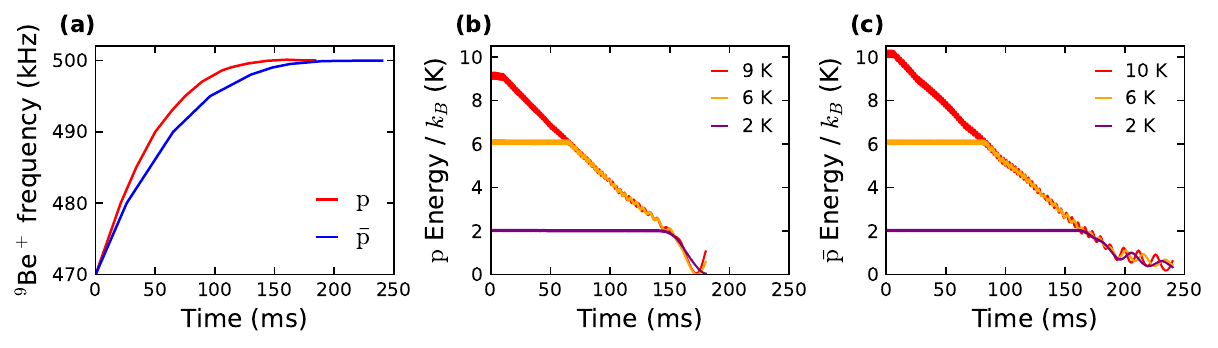}
    \caption{\textbf{Frequency sweep.} \textbf{a)} Frequency values during the sweep of the \be ion potential, which is achieved using a series of linear voltage ramps. The sweep rate is chosen to maintain resonance between the \antiproton and the \be ion throughout the sweep. Examples of the \textbf{b)} proton and \textbf{c)} antiproton energy evolution during the sweep. Voltage fluctuations are not included in these simulations.}
    \label{fig:freqsweep}
\end{figure}

Due to the anharmonicities of the potential, the axial oscillation frequency of the \antiproton decreases with increasing axial energy. To maintain resonance with the \be ion during coupling, we sweep the frequency of the \be potential minimum, as shown in Fig.~\ref{fig:freqsweep}a. The sweep rate is optimized to be as fast as possible while keeping the particles in resonance. We consider $f=$ \SI{500}{\kilo\hertz} and $s_0=$ \SI{0.7}{\milli\meter}, which provide a potential barrier height sufficient to trap more than 90\% of \SI{4}{\kelvin} \mbox{(anti-)protons}. We set \mbox{$\Delta s_0=$ \SI{0}{\micro\meter}}, as maximizing the harmonicity of the \antiproton well is not necessary for anharmonic coupling. Examples of the energy evolution are shown in Fig.~\ref{fig:freqsweep}b for the proton and in Fig.~\ref{fig:freqsweep}c for the antiproton. At the start of the sweep, while remaining off-resonance with the \be ion, the \antiproton energy remains constant; once its oscillation frequency matches that of the \be ion, its energy begins to decrease. A key advantage of the frequency-sweep method is that it does not require prior knowledge of the initial \antiproton energy and the maximum initial energy that can be cooled is set by the depth of the \antiproton potential well. For the parameters considered here, this method enables cooling of any initial energy below \SI{9.4}{\kelvin}$\times k_{\mathrm{B}}$ for protons and below \SI{10.3}{\kelvin}$\times k_{\mathrm{B}}$ for antiprotons. Experimentally, such a sweep could be implemented through a series of voltage ramps. 

\begin{figure}[h]
    \centering
    \includegraphics[width=\textwidth]{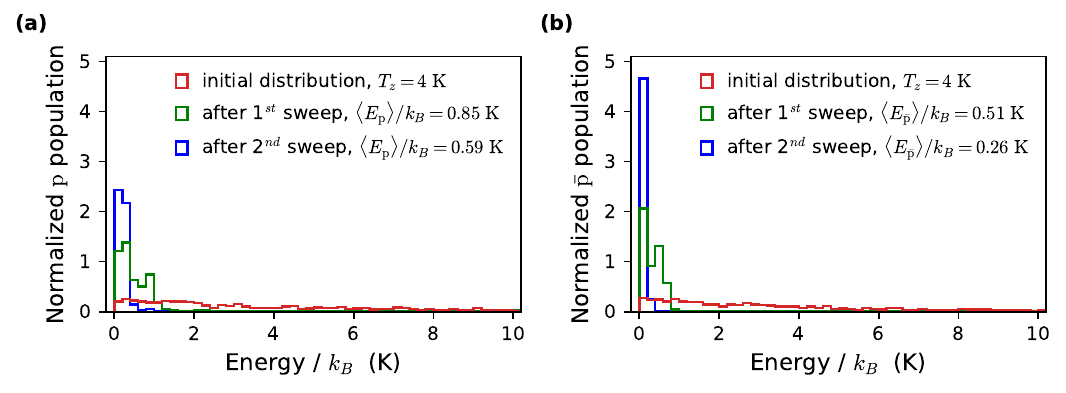}
    \caption{\textbf{Monte Carlo simulation results.} \textbf{(a)} Proton. \textbf{(b)} Antiproton. Energy distributions are shown before and after applying the frequency sweep once and twice. Each plot is based on 1200 Monte Carlo samples. Normally distributed voltage fluctuations with a standard deviation of $\sigma_V = \SI{250}{\nano\volt}$ were applied to all trap electrodes. All distributions are normalized such that the total area under each curve equals one. The \antiproton is initialized with energies drawn from a \SI{4}{\kelvin} Boltzmann distribution. The temperature of the resulting distributions is extracted from the mean energy, as indicated in the legend.}
    \label{fig:montecarlo}
\end{figure}

In Fig.~\ref{fig:montecarlo}, we show the energy distributions of the \antiproton before and after applying the frequency sweep once and twice. We considered 1200 samples, applying voltage fluctuations to the trap electrodes in each case, following a normal distribution with a standard deviation of $\sigma_V = \SI{250}{\nano\volt}$. After two sweeps, the proton energy distribution exhibits a sharp drop at \SI{460}{\milli\kelvin}$\times k_{\rm{B}}$, encompassing 88\% of protons. For the antiproton, the energy distribution exhibits a sharp drop at \SI{260}{\milli\kelvin}$\times k_{\rm{B}}$, encompassing 91\% of antiprotons. Thus, following cooling by frequency sweeping, the majority of the \SI{4}{\kelvin} \antiproton distribution can subsequently be cooled via the harmonic coupling mechanism described in Fig.~\ref{fig:harmoniccoupling}. The resulting energy distributions obtained without including voltage fluctuations are shown in \ref{sec:robustness_estimation}.

\subsection{Discussion}

A robust scheme to cool the \antiproton to its motional ground state could proceed as follows:

\begin{enumerate}
    \item \textbf{Initialization}. Cool the \be ion to its motional ground state and the \antiproton to approximately \SI{4}{\kelvin}.
    \item \textbf{Frequency Sweep}. Perform two frequency sweeps of the \be potential well to cool \SI{91}{\%} (\SI{88}{\%}) of antiprotons (protons) to $<$ \SI{260}{\milli\kelvin} ($<$ \SI{460}{\milli\kelvin}).
    \item \textbf{Harmonic Coupling}. Apply a time-independent harmonic coupling potential to cool the \antiproton to below \SI{1}{\milli\kelvin}.
    \item \textbf{Ground-state Cooling}. Apply a time-independent harmonic coupling potential to bring the \antiproton to its motional ground-state, using a smaller $s_0$ or $f$ than in step (iii) to enhance robustness and decrease $\tau_{ex}$.
    \item \textbf{Verification}. Confirm ground-state cooling by measuring the temperature of the \be ion.
\end{enumerate}

Note that steps (iii) and (iv) can be repeated multiple times, depending on the achievable energy-transfer efficiency. For example, if the transfer efficiency is \SI{80}{\%} then step (iii) should be repeated at least four times, and step (iv) at least thrice. Also note that the \be ion needs to be reinitialized in the motional ground state after each step. The duration of the \antiproton cooling does not impact the time budget for our QLS-based measurements~\cite{Cornejo_2021}, as it serves as an initialization procedure and it does not need to be repeated after the first successful attempt. Using Tab.~\ref{tab:cooling_parameters}, the total time for the ground-state cooling can be estimated. In addition to the reinitialization of the \be ion and the transport of the particles, the total time amounts to approximately \SI{490}{\milli\second} for the proton and \SI{580}{\milli\second} for the antiproton.

\begin{table}[H]
    \centering
    \caption{Possible trapping parameters for each cooling step that bring 88\% of \SI{4}{\kelvin} protons and 91\% antiprotons into the motional ground state. The choice of a different $f_{\rm Be}$ from the particle frequency accounts for anharmonicities or follows from Eq.~\ref{eqn:freq_shift}. The time column provides an estimate of the duration of each step, assuming \SI{80}{\%} of energy transfer at steps (iii) and (iv).}
    \label{tab:cooling_parameters}
    \begin{tabular}{l l l l l l l}
        \toprule
        \toprule
        \multicolumn{6}{c}{\textbf{Proton}} \\
        \midrule
        Cooling step
        & $f_{\rm{p}}$ (kHz)
        & $f_{\rm Be}$ (kHz)
        & $s_0$ (mm)
        & Time (ms)
        & $E_{\rm fin}$ (mK$\times k_{\rm{B}}$) \\
        \midrule
        1) Sweep        & 500 & 470--500 & 0.7 & $2\times180.0$ & $<460$ \\
        2) Harmonic     & 400 & 400.023  & 0.7 & $4\times29.4$  & $<1$   \\
        3) Ground-state & 100 & 100.144  & 0.6 & $3\times4.6$   & Ground state \\
        \midrule
        \midrule
        \multicolumn{6}{c}{\textbf{Antiproton}} \\
        \midrule
        Cooling step
        & $f_{\rm{\bar{p}}}$ (kHz)
        & $f_{\rm Be}$ (kHz)
        & $s_0$ (mm)
        & Time (ms)
        & $E_{\rm fin}$ (mK$\times k_{\rm{B}}$) \\
        \midrule
        1) Sweep        & 500 & 470--500 & 0.7 & $2\times242$   & $<260$ \\
        2) Harmonic     & 450 & 449.968  & 0.6 & $4\times20.8$  & $<1$   \\
        3) Ground-state & 100 & 99.856   & 0.6 & $3\times4.6$   & Ground state \\
        \bottomrule
        \bottomrule
    \end{tabular}
\end{table}

Finally, the axial heating rate in the Coupling trap needs to be taken into account for a successful cooling process. For reference, the cyclotron heating rate in the BASE analysis trap, which has a diameter of $d=$ \SI{3.6}{\milli\meter}, is \SI{6}{quanta\per\hour}~\cite{Borchert2019}. Assuming that anomalous heating in a Penning trap scales as $d^{-2}$ or $d^{-4}$~\cite{Goodwin2016}, we estimate the heating rate in our Coupling trap to be $\scriptstyle{\lesssim}$ \SI{0.7}{quanta\per\second}. Since the final cooling step can be performed within \SI{5}{\milli\second}, this heating rate can be considered negligible. 

\section{Conclusion}

We have simulated the sympathetic cooling dynamics of single ions interacting via the Coulomb force in double-well potentials achievable in current cryogenic multi-Penning traps. This approach is directly relevant to CPT-symmetry tests in the baryonic sector, where we focus on the sympathetic cooling of an \antiproton using a \mbox{ground-state-cooled} \be ion. We showed that the time-independent potential alone is insufficient to efficiently cool \antiprotons initialized at \SI{4}{\kelvin}, as  realistic voltage constraints and anharmonicities hinder sympathetic cooling. To overcome these limitations, we introduced a \mbox{frequency-sweeping} scheme that dynamically maintains resonance between the coupled particles. For the trapping parameters considered in this work, this method enables 88\% of protons to reach energies below \SI{460}{\milli\kelvin}$\times k_{\rm{B}}$ and 91\% of antiprotons to reach energies below \SI{260}{\milli\kelvin}$\times k_{\rm{B}}$. When followed by a time-independent harmonic coupling stage, the \antiproton can be cooled to its motional ground state. This method is applicable to other ion species which cannot be directly laser cooled, such as $\text{H}_2^+$ and $\bar{\text{H}}_2^-$~\cite{Schiller2026}, and provides a viable pathway toward quantum logic spectroscopy of \mbox{(anti-)particles} for high-precision tests of fundamental symmetries in Penning traps.

\ack{}

This work was supported by PTB, LUH, and DFG through the clusters of excellence QUEST and QuantumFrontiers as well as through the Collaborative Research Center SFB1227 (DQ-mat Project-ID 274200144) and ERC StG ``QLEDS''. We also acknowledge financial support from the RIKEN Pioneering Project Funding and the MPG/RIKEN/PTB Center for Time, Constants and Fundamental Symmetries. JMC acknowledges the grant "RYC2023-042535-I" funded by MICIU/AEI/10.13039/501100011033 and by “ESF+”.

\section*{References}

\bibliography{iopart_num, bibliography_qls}

@article{blakestadNeargroundstateTransportTrappedion2011,
  title = {Near-Ground-State Transport of Trapped-Ion Qubits through a Multidimensional Array},
  author = {Blakestad, R. B and Ospelkaus, C. and VanDevender, A. P and Wesenberg, J. H and Biercuk, M. J and Leibfried, D. and Wineland, D. J},
  year = {2011},
  journal = {Physical Review A},
  volume = {84},
  number = {3},
  pages = {032314},
  doi = {10.1103/PhysRevA.84.032314},
  url = {https://doi.org/10.1103/PhysRevA.84.032314},
  urldate = {2011-09-17}
}

@article{bohmanSympatheticCoolingTrapped2021,
  title = {Sympathetic Cooling of a Trapped Proton Mediated by an {{LC}} Circuit},
  author = {Bohman, M. and Grunhofer, V. and Smorra, C. and Wiesinger, M. and Will, C. and Borchert, M. J. and Devlin, J. A. and Erlewein, S. and Fleck, M. and Gavranovic, S. and Harrington, J. and Latacz, B. and Mooser, A. and Popper, D. and Wursten, E. and Blaum, K. and Matsuda, Y. and Ospelkaus, C. and Quint, W. and Walz, J. and Ulmer, S.},
  year = {2021},
  month = aug,
  journal = {Nature},
  volume = {596},
  number = {7873},
  pages = {514--518},
  publisher = {{Nature Publishing Group}},
  issn = {1476-4687},
  doi = {10.1038/s41586-021-03784-w},
  url = {https://doi.org/10.1038/s41586-021-03784-w},
  urldate = {2021-09-16},
  copyright = {2021 The Author(s)},
  langid = {english}
}

@article{brownCoupledQuantizedMechanical2011,
  title = {Coupled Quantized Mechanical Oscillators},
  author = {Brown, K. R. and Ospelkaus, C. and Colombe, Y. and Wilson, A. C. and Leibfried, D. and Wineland, D. J.},
  year = {2011},
  month = mar,
  journal = {Nature},
  volume = {471},
  number = {7337},
  pages = {196--199},
  issn = {1476-4687},
  doi = {10.1038/nature09721},
  url = {https://doi.org/10.1038/nature09721},
  urldate = {2019-01-30},
  copyright = {2011 Nature Publishing Group},
  langid = {english}
}

@article{brownPrecisionSpectroscopyCharged1982,
  title = {Precision Spectroscopy of a Charged Particle in an Imperfect {{Penning}} Trap},
  author = {Brown, Lowell S. and Gabrielse, Gerald},
  year = {1982},
  month = apr,
  journal = {Physical Review A},
  volume = {25},
  number = {4},
  pages = {2423--2425},
  doi = {10.1103/PhysRevA.25.2423},
  url = {https://doi.org/10.1103/PhysRevA.25.2423},
  urldate = {2015-05-19}
}

@article{chouPreparationCoherentManipulation2017,
  title = {Preparation and Coherent Manipulation of Pure Quantum States of a Single Molecular Ion},
  author = {Chou, Chin-wen and Kurz, Christoph and Hume, David B. and Plessow, Philipp N. and Leibrandt, David R. and Leibfried, Dietrich},
  year = {2017},
  month = may,
  journal = {Nature},
  volume = {545},
  number = {7653},
  pages = {203--207},
  issn = {1476-4687},
  doi = {10.1038/nature22338},
  url = {https://doi.org/10.1038/nature22338},
  urldate = {2019-08-28},
  copyright = {2017 Nature Publishing Group},
  langid = {english}
}

@article{ciracQuantumComputationsCold1995,
  title = {Quantum {{Computations}} with {{Cold Trapped Ions}}},
  author = {Cirac, J. I. and Zoller, P.},
  year = {1995},
  month = may,
  journal = {Physical Review Letters},
  volume = {74},
  number = {20},
  pages = {4091},
  doi = {10.1103/PhysRevLett.74.4091},
  url= {https://doi.org/10.1103/PhysRevLett.74.4091},
  urldate = {2011-10-10}
}

@article{cornejoOpticalStimulatedRamanSideband2023,
  title = {Optical Stimulated-{{Raman}} Sideband Spectroscopy of a Single {\textsuperscript{9}}{{Be}}{\textsuperscript{+}} Ion in a {{Penning}} Trap},
  author = {Cornejo, Juan M. and Brombacher, Johannes and Coenders, Julia A. and {von Boehn}, Moritz and Meiners, Teresa and Niemann, Malte and Ulmer, Stefan and Ospelkaus, Christian},
  year = {2023},
  month = sep,
  journal = {Physical Review Research},
  volume = {5},
  number = {3},
  pages = {033226},
  publisher = {{American Physical Society}},
  doi = {10.1103/PhysRevResearch.5.033226},
  url = {https://doi.org/10.1103/PhysRevResearch.5.033226},
  urldate = {2023-09-30}
}

@article{diedrichLaserCoolingZeroPoint1989,
  title = {Laser {{Cooling}} to the {{Zero-Point Energy}} of {{Motion}}},
  author = {Diedrich, F. and Bergquist, J. C. and Itano, Wayne M. and Wineland, D. J.},
  year = {1989},
  month = jan,
  journal = {Physical Review Letters},
  volume = {62},
  number = {4},
  pages = {403},
  doi = {10.1103/PhysRevLett.62.403},
  url = {https://doi.org/10.1103/PhysRevLett.62.403},
  urldate = {2010-09-21},
  lccn = {0000}
}

@article{harlanderTrappedionAntennaeTransmission2011,
  title = {Trapped-Ion Antennae for the Transmission of Quantum Information},
  author = {Harlander, M. and Lechner, R. and Brownnutt, M. and Blatt, R. and Hansel, W.},
  year = {2011},
  month = mar,
  journal = {Nature},
  volume = {471},
  number = {7337},
  pages = {200--203},
  issn = {0028-0836},
  doi = {10.1038/nature09800},
  url = {https://doi.org/10.1038/nature09800},
  urldate = {2011-03-14},
  lccn = {0000}
}

@article{heinzenQuantumlimitedCoolingDetection1990,
  title = {Quantum-Limited Cooling and Detection of Radio-Frequency Oscillations by Laser-Cooled Ions},
  author = {Heinzen, D. J. and Wineland, D. J.},
  year = {1990},
  journal = {Physical Review A},
  volume = {42},
  number = {5},
  pages = {2977},
  doi = {10.1103/PhysRevA.42.2977},
  url = {https://doi.org/10.1103/PhysRevA.42.2977},
  urldate = {2010-09-17},
  lccn = {0000}
}

@article{larsonSympatheticCoolingTrapped1986,
  title = {Sympathetic Cooling of Trapped Ions: {{A}} Laser-Cooled Two-Species Nonneutral Ion Plasma},
  shorttitle = {Sympathetic Cooling of Trapped Ions},
  author = {Larson, D. J. and Bergquist, J. C. and Bollinger, J. J. and Itano, Wayne M. and Wineland, D. J.},
  year = {1986},
  month = jul,
  journal = {Physical Review Letters},
  volume = {57},
  number = {1},
  pages = {70--73},
  doi = {10.1103/PhysRevLett.57.70},
  url = {https://doi.org/10.1103/PhysRevLett.57.70},
  urldate = {2019-01-30}
}

@article{lehnertCPTSymmetryIts2016,
  title = {{{CPT Symmetry}} and {{Its Violation}}},
  author = {Lehnert, Ralf},
  year = {2016},
  month = nov,
  journal = {Symmetry},
  volume = {8},
  number = {11},
  pages = {114},
  publisher = {{Multidisciplinary Digital Publishing Institute}},
  doi = {10.3390/sym8110114},
  url = {https://doi.org/10.3390/sym8110114},
  urldate = {2021-08-09},
  copyright = {http://creativecommons.org/licenses/by/3.0/},
  langid = {english}
}

@article{ludlowOpticalAtomicClocks2015,
  title = {Optical Atomic Clocks},
  author = {Ludlow, Andrew D. and Boyd, Martin M. and Ye, Jun and Peik, E. and Schmidt, P. O.},
  year = {2015},
  month = jun,
  journal = {Reviews of Modern Physics},
  volume = {87},
  number = {2},
  pages = {637--701},
  issn = {0034-6861, 1539-0756},
  doi = {10.1103/RevModPhys.87.637},
  url = {https://doi.org/10.1103/RevModPhys.87.637},
  urldate = {2020-06-09}
}

@article{mickeCoherentLaserSpectroscopy2020,
  title = {Coherent Laser Spectroscopy of Highly Charged Ions Using Quantum Logic},
  author = {Micke, P. and Leopold, T. and King, S. A. and Benkler, E. and Spie{\ss}, L. J. and Schm{\"o}ger, L. and Schwarz, M. and {Crespo L{\'o}pez-Urrutia}, J. R. and Schmidt, P. O.},
  year = {2020},
  month = feb,
  journal = {Nature},
  volume = {578},
  number = {7793},
  pages = {60--65},
  publisher = {{Nature Publishing Group}},
  issn = {1476-4687},
  doi = {10.1038/s41586-020-1959-8},
  url = {https://doi.org/10.1038/s41586-020-1959-8},
  urldate = {2021-07-23},
  copyright = {2020 The Author(s), under exclusive licence to Springer Nature Limited},
  langid = {english}
}

@article{mielke139GHzUV2021,
  title = {139 {{GHz UV}} Phase-Locked {{Raman}} Laser System for Thermometry and Sideband Cooling of {\textsuperscript{9}}{{Be}}{\textsuperscript{+}} Ions in a {{Penning}} Trap},
  author = {Mielke, J. and Pick, J. and Coenders, J. A. and Meiners, T. and Niemann, M. and Cornejo, J. M. and Ulmer, S. and Ospelkaus, C.},
  year = {2021},
  month = oct,
  journal = {Journal of Physics B: Atomic, Molecular and Optical Physics},
  volume = {54},
  number = {19},
  pages = {195402},
  publisher = {{IOP Publishing}},
  issn = {0953-4075},
  doi = {10.1088/1361-6455/ac319d},
  url = {https://doi.org/10.1088/1361-6455/ac319d},
  urldate = {2021-11-21},
  langid = {english}
}

@article{niemannCryogenic9BePenning2019,
  title = {Cryogenic {\textsuperscript{9}}{{Be}}{\textsuperscript{+}} {{Penning}} Trap for Precision Measurements with (Anti-)Protons},
  author = {Niemann, M. and Meiners, T. and Mielke, J. and Borchert, M. J. and Cornejo, J. M. and Ulmer, S. and Ospelkaus, C.},
  year = {2019},
  journal = {Measurement Science and Technology},
  volume = {31},
  number = {3},
  pages = {035003},
  issn = {0957-0233},
  doi = {10.1088/1361-6501/ab5722},
  url = {https://doi.org/10.1088/1361-6501/ab5722},
  urldate = {2020-01-03},
  langid = {english}
}

@article{schneiderDoubletrapMeasurementProton2017,
  title = {Double-Trap Measurement of the Proton Magnetic Moment at 0.3 Parts per Billion Precision},
  author = {Schneider, Georg and Mooser, Andreas and Bohman, Matthew and Sch{\"o}n, Natalie and Harrington, James and Higuchi, Takashi and Nagahama, Hiroki and Sellner, Stefan and Smorra, Christian and Blaum, Klaus and Matsuda, Yasuyuki and Quint, Wolfgang and Walz, Jochen and Ulmer, Stefan},
  year = {2017},
  month = nov,
  journal = {Science},
  volume = {358},
  number = {6366},
  pages = {1081--1084},
  issn = {0036-8075, 1095-9203},
  doi = {10.1126/science.aan0207},
  url = {https://doi.org/10.1126/science.aan0207},
  urldate = {2018-11-29},
  copyright = {Copyright \textcopyright{} 2017, American Association for the Advancement of Science. http://www.sciencemag.org/about/science-licenses-journal-article-reuseThis is an article distributed under the terms of the Science Journals Default License.},
  langid = {english},
  pmid = {29170238}
}

@article{smorraBASEBaryonAntibaryon2015,
  title = {{{BASE}} \textendash{} {{The Baryon Antibaryon Symmetry Experiment}}},
  author = {Smorra, C. and Blaum, K. and Bojtar, L. and Borchert, M. and Franke, K.A. and Higuchi, T. and Leefer, N. and Nagahama, H. and Matsuda, Y. and Mooser, A. and Niemann, M. and Ospelkaus, C. and Quint, W. and Schneider, G. and Sellner, S. and Tanaka, T. and Van Gorp, S. and Walz, J. and Yamazaki, Y. and Ulmer, S.},
  year = {2015},
  month = nov,
  journal = {The European Physical Journal Special Topics},
  volume = {224},
  number = {16},
  pages = {3055},
  issn = {1951-6355, 1951-6401},
  doi = {10.1140/epjst/e2015-02607-4},
  url = {https://doi.org/10.1140/epjst/e2015-02607-4},
  urldate = {2015-11-23},
  langid = {english}
}

@article{winelandExperimentalIssuesCoherent1998,
  title = {Experimental Issues in Coherent Quantum-State Manipulation of Trapped Atomic Ions},
  author = {Wineland, D.J. and Monroe, C. and Itano, W.M. and Leibfried, D. and King, B.E. and Meekhof, D.M.},
  year = {1998},
  month = may,
  journal = {Journal of Research of the National Institute of Standards and Technology},
  volume = {103},
  number = {3},
  pages = {259},
  issn = {1044677X},
  doi = {10.6028/jres.103.019},
  url = {https://doi.org/10.6028/jres.103.019},
  urldate = {2019-12-13},
  langid = {english}
}

@article{wolfNondestructiveStateDetection2016,
  title = {Non-Destructive State Detection for Quantum Logic Spectroscopy of Molecular Ions},
  author = {Wolf, Fabian and Wan, Yong and Heip, Jan C. and Gebert, Florian and Shi, Chunyan and Schmidt, Piet O.},
  year = {2016},
  month = feb,
  journal = {Nature},
  volume = {530},
  number = {7591},
  pages = {457--460},
  issn = {1476-4687},
  doi = {10.1038/nature16513},
  url = {https://doi.org/10.1038/nature16513},
  urldate = {2019-08-28},
  copyright = {2016 Nature Publishing Group},
  langid = {english}
}

@article{cornejo_resolved-sideband_2024,
	title = {Resolved-sideband cooling of a single {Be} + 9 ion in a cryogenic multi-{Penning}-trap for discrete symmetry tests with (anti-)protons},
	volume = {6},
	issn = {2643-1564},
	url = {https://link.aps.org/doi/10.1103/PhysRevResearch.6.033233},
	doi = {10.1103/PhysRevResearch.6.033233},
	language = {en},
	number = {3},
	urldate = {2024-09-06},
	journal = {Physical Review Research},
	author = {Cornejo, Juan M. and Brombacher, Johannes and Coenders, Julia A. and Von Boehn, Moritz and Meiners, Teresa and Niemann, Malte and Ulmer, Stefan and Ospelkaus, Christian},
	month = sep,
	year = {2024},
	pages = {033233},
}

@phdthesis{Wiesinger2023,
  author       = {Wiesinger, Markus},
  title        = {Sympathetic Cooling of a Single Individually-Trapped Proton in a Cryogenic Penning Trap},
  school       = {Ruprecht-Karls Universität Heidelberg},
  year         = {2023},
  type         = {{P}h{D} Thesis},
  url          = {https://hdl.handle.net/21.11116/0000-000D-A678-1},
  note         = {{MPI} for Nuclear Physics, Max Planck Society}  
}

@manual{stahl_manual,
  title        = {UM Series Installation and User Manual, Rev.\ 2025c},
  author       = {Stahl Electronics},
  year         = {2025},
  month        = {January 31},
  note         = {UM 1-14 (Options -LN, -SW), UM 1-32 LN, UM 1-60 LN-SW; see p.\ 17},
  url          = {https://www.stahl-electronics.com/devices/um/Manual_UM_LN_SW_V2025c.pdf},
  organization = {Stahl Electronics, Mettenheim, Germany}
}

@unpublished{Schiller2026,
  author = {Schiller, Stefan and Cornejo, Juan M. and Poljakov, Nikita and Ospelkaus, Christian and Ulmer, Stefan and Bakalov, Dimitar},
  note   = {submitted to \emph{Mol. Phys.}},
  year   = {2026}
}

@unpublished{Julia2026,
  author       = {Coenders, J. A. and others},
  note         = {Manuscript in preparation}
}

@article{Will2024,
  title = {Image-Current Mediated Sympathetic Laser Cooling of a Single Proton in a Penning Trap Down to 170 mK Axial Temperature},
  author = {Will, C. and others},
  collaboration = {BASE Collaboration},
  journal = {Phys. Rev. Lett.},
  volume = {133},
  issue = {2},
  pages = {023002},
  numpages = {6},
  year = {2024},
  month = {Jul},
  publisher = {American Physical Society},
  doi = {10.1103/PhysRevLett.133.023002},
  url = {https://doi.org/10.1103/PhysRevLett.133.023002}
}

@article{Nagahama2016,
    author = {Nagahama, H. and others},
    title = {Highly sensitive superconducting circuits at ∼700 kHz with tunable quality factors for image-current detection of single trapped antiprotons},
    journal = {Review of Scientific Instruments},
    volume = {87},
    number = {11},
    pages = {113305},
    year = {2016},
    month = {11},
    issn = {0034-6748},
    doi = {10.1063/1.4967493},
    url = {https://doi.org/10.1063/1.4967493}
}

@article{Nitzschke2020,
  author       = {Nitzschke, Diana and others},
  title        = {Elementary laser-less quantum logic operations with (anti-)protons in Penning traps},
  journal      = {Advanced Quantum Technologies},
  volume       = {3},
  number       = {11},
  pages        = {1900133},
  year         = {2020},
  doi          = {10.1002/qute.201900133},
  url          = {https://doi.org/10.1002/qute.201900133}
}

@article{Brown1986,
  title = {Geonium theory: Physics of a single electron or ion in a Penning trap},
  author = {Lowell S. Brown and Gerald Gabrielse},
  journal = {Rev. Mod. Phys.},
  volume = {58},
  number = {1},
  pages = {233--311},
  year = {1986},
  doi = {10.1103/RevModPhys.58.233},
  publisher = {American Physical Society},
  url={https://doi.org/10.1103/RevModPhys.58.233}
}

@article{Powell2003,
  author    = {Powell, H F and Echaniz, S R de and Phillips, E S and Segal, D M and Thompson, R C},
  journal   = {Journal of Physics B: Atomic, Molecular and Optical Physics},
  title     = {Improvement of laser cooling of ions in a Penning trap by use of the axialization technique},
  year      = {2003},
  issn      = {0953-4075},
  month     = feb,
  number    = {5},
  pages     = {961--970},
  volume    = {36},
  doi       = {10.1088/0953-4075/36/5/315},
  url = {https://doi.org/10.1088/0953-4075/36/5/315},
  publisher = {IOP Publishing},
}

@article{cornell1990,
  title = {Mode coupling in a Penning trap: \ensuremath{\pi} pulses and a classical avoided crossing},
  author = {Cornell, Eric A. and others},
  journal = {Phys. Rev. A},
  volume = {41},
  issue = {1},
  pages = {312--315},
  numpages = {0},
  year = {1990},
  month = {Jan},
  publisher = {American Physical Society},
  doi = {10.1103/PhysRevA.41.312},
  url = {https://doi.org/10.1103/PhysRevA.41.312}
}

@article{verlet1967,
  author    = {Verlet, Loup},
  title     = {Computer 'Experiments' on Classical Fluids. I. Thermodynamical Properties of Lennard-Jones Molecules},
  journal   = {Phys. Rev.},
  volume    = {159},
  pages     = {98--103},
  year      = {1967},
  doi       = {10.1103/PhysRev.159.98},
  url       = {https://doi.org/10.1103/PhysRev.159.98}
}

@article{penrose1955,
  author    = {Penrose, Roger},
  title     = {A generalized inverse for matrices},
  journal   = {Math. Proc. Cambridge Philos. Soc.},
  volume    = {51},
  number    = {3},
  pages     = {406--413},
  year      = {1955},
  doi       = {10.1017/S0305004100030401},
  url       ={https://doi.org/10.1017/S0305004100030401}
}

@article{Latacz2025,
author={Latacz, B. M.
and others},
title={Coherent spectroscopy with a single antiproton spin},
journal={Nature},
year={2025},
month={Jul},
day={23},
issn={1476-4687},
doi={10.1038/s41586-025-09323-1},
url={https://doi.org/10.1038/s41586-025-09323-1}
}

@article{Leonhardt2025,
author={Leonhardt, M.
and others},
title={Proton transport from the antimatter factory of CERN},
journal={Nature},
year={2025},
month={May},
day={01},
volume={641},
number={8064},
pages={871-875},
issn={1476-4687},
doi={10.1038/s41586-025-08926-y},
url={https://doi.org/10.1038/s41586-025-08926-y}
}

@article{Goodwin2016,
  title = {Resolved-Sideband Laser Cooling in a Penning Trap},
  author = {Goodwin, J. F. and others},
  journal = {Phys. Rev. Lett.},
  volume = {116},
  issue = {14},
  pages = {143002},
  numpages = {5},
  year = {2016},
  month = {Apr},
  publisher = {American Physical Society},
  doi = {10.1103/PhysRevLett.116.143002},
  url = {https://doi.org/10.1103/PhysRevLett.116.143002}
}

@article{Borchert2019,
  title = {Measurement of Ultralow Heating Rates of a Single Antiproton in a Cryogenic Penning Trap},
  author = {Borchert, M. J. and others},
  journal = {Phys. Rev. Lett.},
  volume = {122},
  issue = {4},
  pages = {043201},
  numpages = {5},
  year = {2019},
  month = {Jan},
  publisher = {American Physical Society},
  doi = {10.1103/PhysRevLett.122.043201},
  url = {https://doi.org/10.1103/PhysRevLett.122.043201}
}

@misc{comsol,
  title = {{COMSOL Multiphysics\textsuperscript{\tiny\textregistered}}},
  year = {2025},
  url = {https://comsol.com/},
  note = {Version 6.2. COMSOL AB, Stockholm, Sweden.}
}

@article{Smorra2017,
author={Smorra, C.
and Sellner, S.
and Borchert, M. J.
and others},
title={A parts-per-billion measurement of the antiproton magnetic moment},
journal={Nature},
year={2017},
month={Oct},
day={01},
volume={550},
number={7676},
pages={371-374},
issn={1476-4687},
doi={10.1038/nature24048},
url={https://doi.org/10.1038/nature24048}
}

@article{Cornejo_2021,
doi = {10.1088/1367-2630/ac136e},
year = {2021},
month = {jul},
publisher = {IOP Publishing},
volume = {23},
number = {7},
pages = {073045},
author = {Juan M Cornejo and Ralf Lehnert and Malte Niemann and others},
title = {Quantum logic inspired techniques for spacetime-symmetry tests with (anti-)protons},
journal = {New J. Phys.},
url={https://doi.org/10.1088/1367-2630/ac136e}
}

@article{schmidt_2005,
author = {P. O. Schmidt and T. Rosenband and C. Langer and others},
title = {Spectroscopy Using Quantum Logic},
journal = {Science},
volume = {309},
number = {5735},
pages = {749-752},
year = {2005},
doi = {10.1126/science.1114375},
url={https://doi.org/10.1126/science.1114375}}

@article{VELARDI201420,
title = {Proton extraction by laser ablation of transition metals},
journal = {Nucl. Instrum. Methods Phys. Res. B},
volume = {331},
pages = {20-22},
year = {2014},
note = {11th European Conference on Accelerators in Applied Research and Technology},
issn = {0168-583X},
doi = {https://doi.org/10.1016/j.nimb.2013.11.029},
url = {https://www.sciencedirect.com/science/article/pii/S0168583X1400113X},
author = {L. Velardi and D. {Delle Side} and J. Krasa and V. Nassisi}}

@article{vBoehn2025,
author={von Boehn, Moritz
and Schaper, Jan
and Coenders, Julia A and others},
title={Speeding up adiabatic ion transport in macroscopic multi-Penning-trap stacks for high-precision experiments},
journal={Commun. Phys.},
year={2025},
month={Mar},
day={19},
volume={8},
number={1},
pages={107},
issn={2399-3650},
doi={10.1038/s42005-025-02031-2},
url={https://doi.org/10.1038/s42005-025-02031-2}
}

@article{Meiners2024,
author={Meiners, Teresa
and Coenders, Julia A.
and Brombacher, Johannes
and others},
title={Fast adiabatic transport of single laser-cooled 9Be+ ions in a cryogenic Penning trap stack},
journal={Eur. Phys. J. Plus},
year={2024},
month={Mar},
day={15},
volume={139},
number={3},
pages={262},
issn={2190-5444},
doi={10.1140/epjp/s13360-024-04936-3},
url={https://doi.org/10.48550/arXiv.2309.06776}}

@article{Borchert2022,
author={Borchert, M. J.
and Devlin, J. A.
and Erlewein, S. R.
and others},
title={A 16-parts-per-trillion measurement of the antiproton-to-proton charge--mass ratio},
journal={Nature},
year={2022},
month={Jan},
day={01},
volume={601},
number={7891},
pages={53-57},
issn={1476-4687},
doi={10.1038/s41586-021-04203-w},
url={https://doi.org/10.1038/s41586-021-04203-w}}

@article{Marshall2025,
  title = {High-Stability Single-Ion Clock with $5.5\ifmmode\times\else\texttimes\fi{}{10}^{\ensuremath{-}19}$ Systematic Uncertainty},
  author = {Marshall, Mason C. and Castillo, Daniel A. Rodriguez and Arthur-Dworschack, Willa J. and Aeppli, Alexander and Kim, Kyungtae and Lee, Dahyeon and Warfield, William and Hinrichs, Joost and Nardelli, Nicholas V. and Fortier, Tara M. and Ye, Jun and Leibrandt, David R. and Hume, David B.},
  journal = {Phys. Rev. Lett.},
  volume = {135},
  issue = {3},
  pages = {033201},
  numpages = {6},
  year = {2025},
  month = {Jul},
  publisher = {American Physical Society},
  doi = {10.1103/hb3c-dk28},
  url = {https://doi.org/10.1103/hb3c-dk28}
}

\appendix
\section{Frequency shift due to the Coulomb interaction}
\label{sec:freq_shift}

Here, we derive Eq.~\ref{eqn:freq_shift}.  
In the absence of Coulomb coupling, let $\omega_a$ denote the axial oscillation frequency of particle~$a$, and let $z_a$ and $z_b$ represent the displacements of particles~$a$ and~$b$ from their respective trap minima along the trap axis. The effective potential $U_{\mathrm{eff}}$ experienced by particle~$a$ is given by the sum of the harmonic trapping potential $U_{\mathrm{trap}}$ and the Coulomb interaction potential $U_C$ due to particle~$b$:
\begin{equation}
    U_{\mathrm{eff}}(z_a, z_b)
    = U_{\mathrm{trap}}(z_a) + U_C(z_a, z_b)
    = \frac{1}{2} m_a \omega_a^2 z_a^2 
    + \frac{1}{4\pi\epsilon_0} \frac{q_a q_b}{s_0 - z_a + z_b}.
\end{equation}
For small displacements $z_a, z_b \ll s_0$, the Coulomb potential can be expanded in a Taylor series around $z_a = z_b = 0$ up to second order:
\begin{equation}
    U_C(z_a, z_b) \approx \frac{q_a q_b}{4\pi\epsilon_0 s_0}
    \left[
        1 + \frac{z_a - z_b}{s_0}
        + \frac{(z_a - z_b)^2}{s_0^2}
    \right].
\end{equation}
The constant term can be omitted as it does not affect the particle dynamics.  
The linear term corresponds to a uniform force that merely shifts the equilibrium positions and is likewise neglected.  
The quadratic term, however, modifies the curvature of the potential and therefore contributes to a shift in the oscillation frequency. The effective potential can thus be written as
\begin{equation}
    U_{\mathrm{eff}} \left(z_a, z_b \right) \approx
    \frac{1}{2} m_a \omega_a^2 z_a^2
    + \frac{q_a q_b}{4\pi\epsilon_0 s_0^3} (z_a - z_b)^2.
    \label{eqn:U_effective}
\end{equation}
Assuming that $U_{\mathrm{eff}}$ remains  harmonic in the vicinity of its minimum, the modified oscillation frequency $\omega_a'$ of particle~$a$ in the presence of the Coulomb interaction is obtained from
\begin{equation}
    \omega_a'^2 = \frac{1}{m_a} \frac{\partial^2 U_{\mathrm{eff}}}{\partial z_a^2}.
\end{equation}
Taking the second derivative of equation~\ref{eqn:U_effective} with respect to $z_a$ and dividing by $m_a$ yields
\begin{equation}
    \omega_a'^2 = \omega_a^2 + \frac{2 q_a q_b}{4\pi\epsilon_0 s_0^3 m_a}.
    \label{eqn:omega_eff}
\end{equation}
Assuming that the Coulomb interaction induces only a small perturbation, such that $\omega_a' \approx \omega_a$, the frequency shift $\Delta \omega_a = \omega_a' - \omega_a$ can be approximated as
\begin{equation}
    \Delta \omega_a 
    = \frac{\omega_a'^2 - \omega_a^2}{2\omega_a}
    \approx \frac{q_a q_b}{4\pi\epsilon_0 s_0^3 m_a \omega_a}.
\end{equation}

Performing an analogous calculation for particle~$b$ yields its corresponding frequency shift $\Delta \omega_b$.  
The relative frequency shift between the two particles is therefore
\begin{equation}
    \Delta \omega = \Delta \omega_a - \Delta \omega_b 
    = \left( \frac{1}{m_a} - \frac{1}{m_b} \right)
      \frac{q_a q_b}{4\pi\epsilon_0 s_0^3 \omega},
\end{equation}
where $\omega \approx \omega_a \approx \omega_b$ has been assumed.

\section{Robustness estimation}
\label{sec:robustness_estimation}

\begin{figure}[h]
    \centering
    \includegraphics[width=\linewidth]{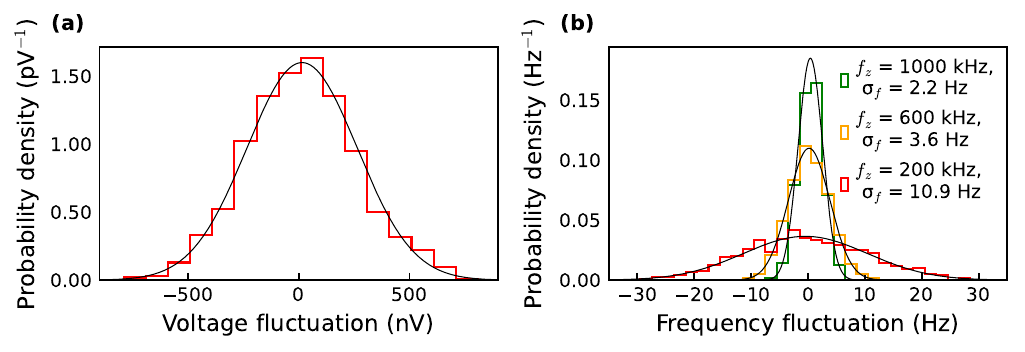}
    \caption{\textbf{Frequency fluctuations
    due to voltage fluctuations.} \textbf{(a)} Voltage fluctuations following a normal distribution with a standard deviation $\sigma_V=$ \SI{250}{\nano\volt}. \textbf{(b)} Examples of resulting frequency fluctuations for a particle separation of $s_0=$ \SI{0.7}{\milli\meter} and different trap frequencies $f_z$. The trap frequency fluctuation are estimated from the difference between the \antiproton and the \be ion at their respective potential minima. The standard deviations, $\sigma_f$, of the normal fits are shown in the legend. Each estimation of $\sigma_f$ is based on a Monte Carlo simulation with 1000 samples. The data are normalized such that the area under each curve equals 1.} 
    \label{fig:gaussian}
\end{figure}

\begin{figure}[h]
    \centering
    \includegraphics[width=\linewidth]{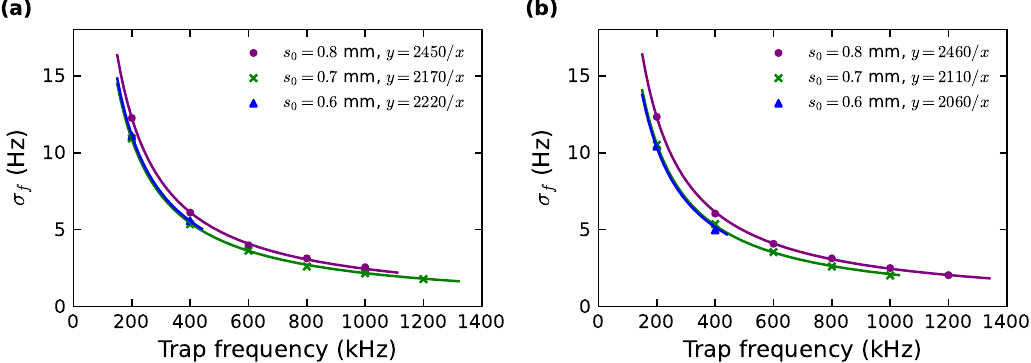}
    \caption{\textbf{Frequency fluctuation results.} \textbf{a)} Proton \textbf{b)} Antiproton. The standard deviations, $\sigma_f$, obtained from the Monte Carlo simulations as described in Fig.~\ref{fig:gaussian}, are shown as a function of the trap frequency. The data is fit to a reciprocal function, as indicated in the legend, with coefficients in \SI{}{\kilo\hertz\times\hertz}.} 
    \label{fig:freqfluctuations}
\end{figure}

To calculate the energy transfer stability shown in Fig.~\ref{fig:stability}, we used the following approach. As illustrated in Figs.~\ref{fig:gaussian} and ~\ref{fig:freqfluctuations}, we estimated the standard deviation $\sigma_f$ of the frequency fluctuations, assuming they follow a normal distribution. These fluctuations arise from voltage noise applied to all trap electrodes, modeled as normally distributed with a standard deviation of $\sigma_V = \SI{250}{\nano\volt}$ for voltages of less than \SI{10}{\volt}. For each data point, we considered 1000 samples of voltage fluctuations. To evaluate the resulting frequency variations, we computed the difference in the oscillation frequencies of the \antiproton and the \be ion at their respective potential minima.

To quantify the robustness of the energy transfer, we estimate the value $\gamma /(2\sigma_f)$, which describes how close the coupling is to the resonance for the resonance width $\gamma=\frac{1}{2\tau_{ex}}$. Therefore, from the Lorentzian equation, we obtain that the robustness sets the fraction of energy $p$ that can be transferred with a $95.4\%$ confidence, given by the following equation,

\begin{equation}
    \frac{\gamma}{2\sigma_f} = \frac{1}{\sqrt{\frac{1}{p}-1}}.
\label{eqn:robustness}
\end{equation}

\begin{figure}[H]
    \centering
    \includegraphics[width=\linewidth]{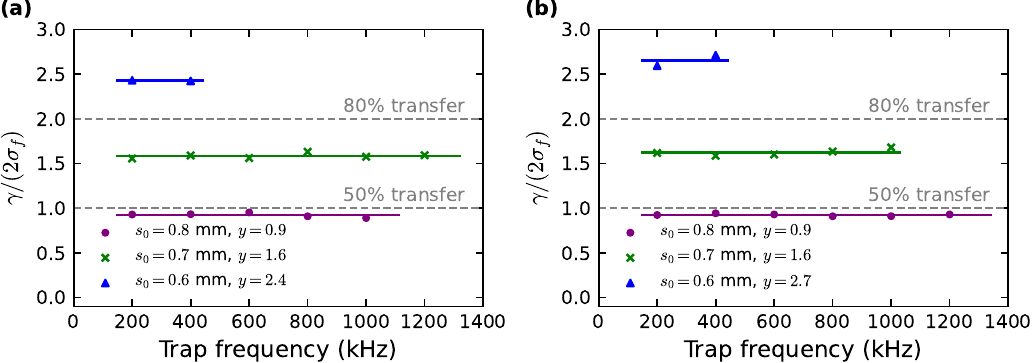}
    \caption{\textbf{Robustness of the energy transfer.} \textbf{a)} Proton. \textbf{b)} Antiproton. We quantify robustness by the ratio of the half width at half maximum, $\gamma$, from Fig.~\ref{fig:stability}a, to two standard deviations in frequency fluctuations, $\sigma_f$, from Fig.~\ref{fig:freqfluctuations}. The ratio $\gamma/(2\sigma_f)$ determines the fraction $p$ of the energy that can be transferred with a $95.4\%$ confidence. The data are fit to a constant function, as indicated in the legend, with unitless coefficients. Dashed gray lines illustrate different robustness values corresponding to different $p$.} 
    \label{fig:robustness}
\end{figure}

We show the robustness as a function of the trap frequency in Fig.~\ref{fig:robustness}. Since the robustness is constant with respect to the trap frequency, we only consider the maximum initial energy $E_{max}$ for each $s_0$ to describe the stability of the transfer. From Fig.~\ref{fig:harmoniccoupling}, we find a corresponding frequency $f_{E_{max}}$ for each $E_{max}$. From Fig.~\ref{fig:freqfluctuations}, we determine the frequency fluctuation $\sigma_{f,{\SI{250}{\nano\volt}, E_{max}}}$ corresponding to $f_{E_{max}}$. From Eq.~\ref{eqn:robustness} we find the maximum allowed frequency fluctuation $\sigma_{f,p=0.8}$ for a $p=0.8$ transfer. To determine the required voltage stability of the power supply, we assume that the frequency fluctuation $\sigma_f$ scales linearly with the voltage fluctuation $\sigma_V$. The corresponding peak-to-peak voltage stability $V_{pp}$ can then be expressed as

\begin{equation}
    V_{pp} = 4 \times \SI{250}{\nano\volt} \times \frac{\sigma_{f,p=0.8}}{\sigma_{f,{\SI{250}{\nano\volt}, E_{max}}}},
    \label{eqn:stability_estimate}
\end{equation}
where the factor of 4 accounts for two conversions: a factor of 2 from the standard deviation $\sigma_V$ to the peak-to-peak voltage $V_{pp}$, and another factor of 2 to convert from the 95.4\% to the 68.3\% confidence interval. The resulting power supply stability is shown in Fig.~\ref{fig:stability}b.

\begin{figure}[H]
    \centering
    \includegraphics[width=\linewidth]{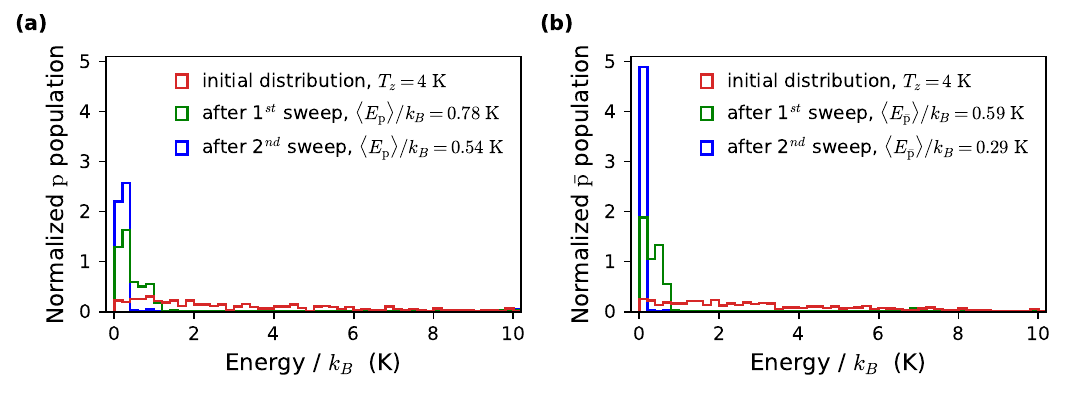}
    \caption{\textbf{Monte Carlo simulation results without voltage noise.} \textbf{a)} Proton. \textbf{b)} Antiproton. Both plots show results for 400 samples. This figure is analogous to Fig.~\ref{fig:montecarlo}, but without voltage noise applied to the trap electrodes.} 
    \label{fig:montecarlo_nonoise}
\end{figure}

Fig.~\ref{fig:montecarlo_nonoise} shows the results of the frequency-sweep Monte Carlo simulations in the absence of voltage noise.  After two sweeps, the proton energy distribution exhibits a sharp drop at \SI{340}{\milli\kelvin}$\times k_{\rm{B}}$, encompassing 89\% of protons. For the antiproton, the energy distribution exhibits a sharp drop at \SI{160}{\milli\kelvin}$\times k_{\rm{B}}$, encompassing 91\% of antiprotons. Therefore, compared to the case including voltage noise in Fig.~\ref{fig:montecarlo}, the resulting distributions are similar, indicating that the frequency-sweep technique is robust against voltage fluctuations.

\end{document}